\documentclass[a4paper,12pt]{article}
\usepackage{algorithm}
\usepackage{algorithmic}
\usepackage{multicol}
\usepackage{multirow}
\usepackage{graphicx}
\usepackage{grffile}
\usepackage{amssymb}
\usepackage{amsmath}

\newcommand\T{\rule{0pt}{2.7ex}}
\newcommand\B{\rule[-.7ex]{5pt}{0pt}}

\newenvironment{nstabbing}
  {\setlength{\topsep}{0pt}%
   \setlength{\partopsep}{0pt}%
   \tabbing}
  {\endtabbing}

\begin{document}

\title{A Multi-phase Approach for Improving Information Diffusion in Social Networks
\thanks{
Please cite the original publication that will be appearing in the Proceedings of The 14th International Conference on Autonomous Agents \& Multiagent Systems, 2015.
This work is funded by Adobe Research Labs, Bangalore, India. The first and second authors are supported by IBM and TCS Doctoral Fellowships, respectively.
The authors thank Surabhi Akotiya for the useful discussions.
}
}

\author{Swapnil Dhamal, Prabuchandran K.J., and Y. Narahari\\
\normalsize Indian Institute of Science, Bangalore, India}
\date{}


\maketitle

\begin{abstract}
For maximizing influence spread in a social network, given a certain budget on the number of seed nodes, we investigate the effects of selecting and activating the seed nodes in multiple phases. In particular, we formulate an appropriate objective function for two-phase influence maximization under the independent cascade model, investigate its properties, and propose algorithms for determining the seed nodes in the two phases. We also study the problem of determining an optimal budget-split and delay between the two phases.
\end{abstract}



\section{Introduction}
Social networks play a fundamental role in the spread of influence on a large scale; this is harnessed by companies for viral marketing.
The problem of  {\em influence maximization} deals with selecting $k$ seed nodes where the diffusion should be triggered,
so as to maximize the influence when diffusion concludes;
we call $k$ as the budget.
This problem
has been extensively studied in the literature
\cite{guille2013information},
including that of AAMAS 
\cite{li2014cross,maghami2012identifying}.
The basic idea of using multiple phases for maximizing an objective function has been presented in \cite{golovin2011adaptive}.
To the best of our knowledge, ours is the first detailed effort to study multi-phase diffusion in social networks.

An advantage of multi-phase diffusion is that the seed nodes in any phase, except the first one, can be chosen based on the spread observed so far, thus having more certainty during seed selection.
But owing to delayed seed selection, the diffusion may be slower, leading to compromise of time.

\section{Problem Formulation}
\label{sec:problem}

As a starting point, 
we focus on two-phase diffusion. 
Given a graph $G$, we consider Independent Cascade (IC) model where, $p_{uv}$ is the probability with which node $u$ can influence $v$. 
Let $X$ be a live graph (got by independently sampling edges in $G$)
 and $p(X)$ be the probability of its occurrence.
Let $\sigma ^X (S)$ be the number of nodes reachable from set $S$ in $X$
(so expected number of influenced nodes at the end of single phase diffusion with seed set $S$ is
$\sigma (S) = \sum_X p(X) \sigma ^X (S)$).

At the beginning (time 0), let $k_1$ seed nodes be selected for first phase and after delay $d$, $k_2$ ($\leq k-k_1$) for second phase. 
We aim to maximize the expected influence at the end of two-phase diffusion. 
For now, assume $k_1,k_2$, $d$ to be given; our objective is to determine seeds for the two phases.

Let $S_1$ be the seed set for first phase
and $X$ be the destined live graph (unknown at time 0).
Let $Y$ be the observed diffusion at time $d$, which gives $\mathcal{A}^Y$ and $\mathcal{R}^Y$, the sets of already and recently influenced nodes, respectively. 
At time $d$, 
given that nodes in $\mathcal{R}^Y$ effectively are seeds for second phase (as per IC model),
we aim to select an additional seed set $S_2 ^{O(Y,k_2)}$ of size $k_2$, that maximizes the final influence.
We obtain $S_2 ^{O(X,S_1,d,k_2)}$
since $Y$ is unique for a particular $(X,S_1,d)$. 
So our objective is to find $S_1$ that maximizes
%

\begin{align*}
g(S_1) 
& =
\sum_{Y} p(Y)  \big\{ |\mathcal{A}^{Y}| + \sum_{X} p(X|Y) \sigma^{X \setminus \mathcal{A}^Y}(\mathcal{R}^Y \cup S_2 ^{O(Y,k_2)})   \big\}
\\ &=
\sum_X p(X) \sigma^X (S_1 \cup S_2 ^{O(X,S_1,d,k_2)}) 
\end{align*}

Note that the choice of $S_2 ^{O(Y,k_2)}=S_2 ^{O(X,S_1,d,k_2)}$ depends not just on $X$, but
on $Y$, and hence on all live graphs that could result from $Y$ (like in single phase, choice of the best seed set depends on all live graphs that could result from $G$).
NP-hardness of maximizing $g(\cdot)$ is clear.
It can be shown that, for fixed $k_2$ and $d$, $g(\cdot)$ is non-negative and monotone increasing 
(note that with $k_2$ and $d$ as variables, $g(\cdot)$ is not monotone), but it is neither submodular nor supermodular.
However, it was observed using simulations on the test graphs, that the diminishing marginal returns property (characteristic of submodular functions) holds in most cases.

\textit{An example for computing $g(\cdot)$}:
A graph with $\{A,B,C,D\}$ as nodes, $p_{AB}=0.5,p_{BC}=0.8,p_{BD}=0.9$.
Consider $S_1=\{A\}$, $k_2=1$, $d=1$.
Table~\ref{tab:example} lists the two possibilities of Y
($S_2^{O(Y,k_2)}$ is easy to compute).
We get $g(\{A\})= 
3.80$.

\begin{table}[h]
\begin{center}
 \begin{tabular}{|c|c|c|c|c|c|}
 \hline 
 \multicolumn{6}{|c|}{\T \B
  $S_1=\{A\}, k_2=1, d=1$} \\
 \hline \T \B \hspace{-.3cm}
   \multirow{2}{*}{$X$} & \multirow{2}{*}{$p(X)$} & \multicolumn{2}{c|}{\T \B $Y$} & \multirow{2}{*}{$S_2^{O(Y,1)}$}  & \multirow{2}{*}{$g(S_1)$}  \\ 
   \cline{3-4}
   \T \B
  & & $\mathcal{A}^Y$ & $\mathcal{R}^Y$  & & \\
    \hline \hline \T \B  \hspace{-.3cm}
    $\{AB,BC,BD\}$ & 0.36 & \multirow{4}{*}{$\{A\}$}  & \multirow{4}{*}{$\{B\}$} &\multirow{4}{*}{$\{C\}$}   & 4\\
    \cline{1-2}\cline{6-6} \hspace{-.3cm} \T \B
   $\{AB,BC\}$  & 0.04 &  & & & 3 \\
    \cline{1-2}\cline{6-6} \hspace{-.3cm} \T \B
      $\{AB,BD\}$ & 0.09 & & &  & 4\\
    \cline{1-2}\cline{6-6} \hspace{-.3cm} \T \B
      $\{AB\}$ & 0.01 &  & &  & 3\\
   \hline \T \B \hspace{-.3cm}
   $\{BC,BD\}$ &  0.36 & \multirow{4}{*}{$\{A\}$} & \multirow{4}{*}{$\{\}$} & \multirow{4}{*}{$\{B\}$} 
   &  4 \\
   \cline{1-2}\cline{6-6} \hspace{-.3cm} \T \B
    $\{BC\}$ & 0.04  &  &  & & 3\\
    \cline{1-2}\cline{6-6} \hspace{-.3cm} \T \B
    $\{BD\}$ & 0.09  & &  & & 3\\
    \cline{1-2}\cline{6-6} \hspace{-.3cm} \T \B
    $\{\}$ & 0.01  & &  & & 2\\ 
    \hline 
 \end{tabular}
 \caption{Table for the example}
 \label{tab:example}
\end{center}
\end{table}

  \begin{figure}
\begin{minipage}{0.5\textwidth}
 \includegraphics[scale=.55]{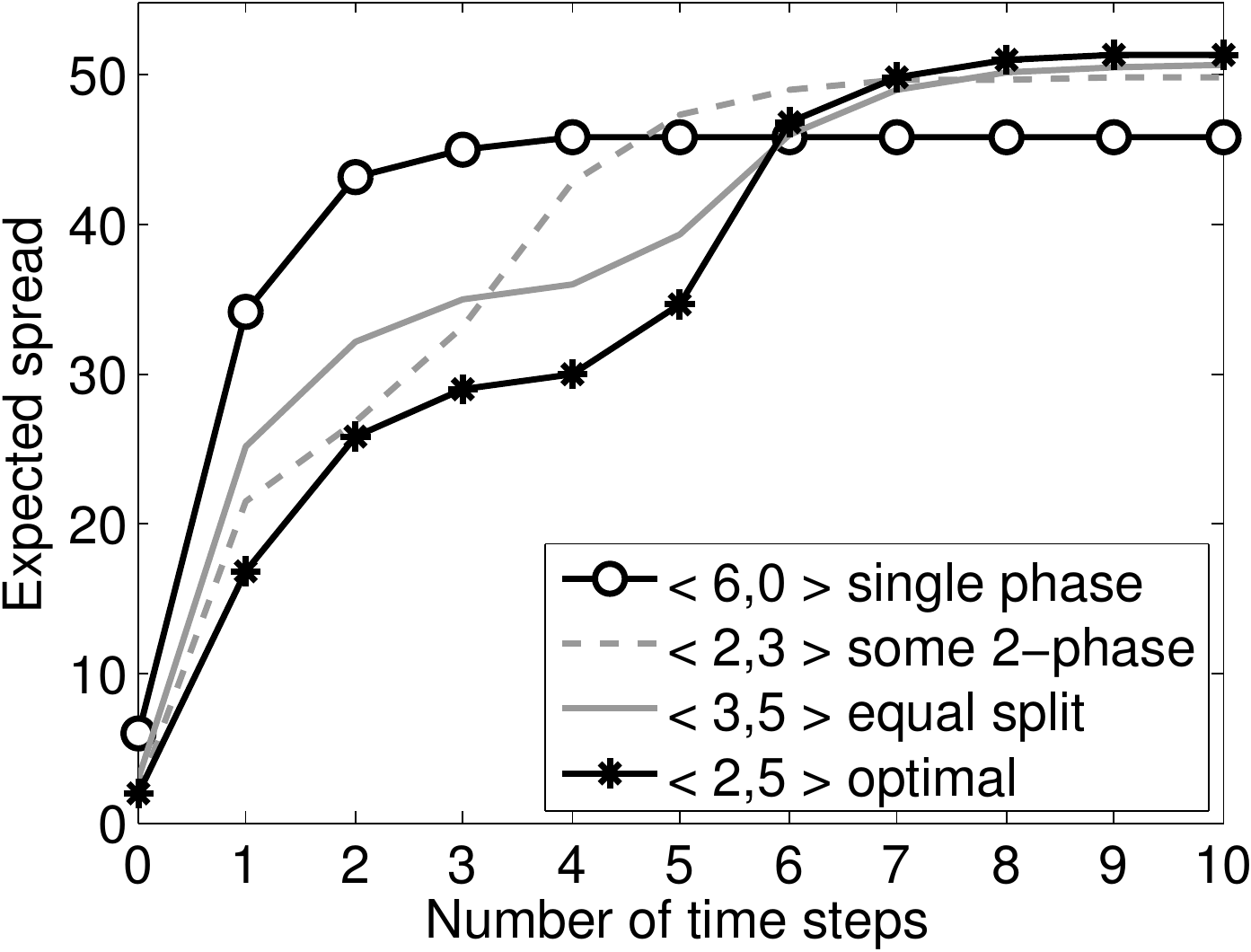} 
  \\ \centering (a)
\end{minipage}
 \begin{minipage}{0.5\textwidth}
  \includegraphics[scale=.55]{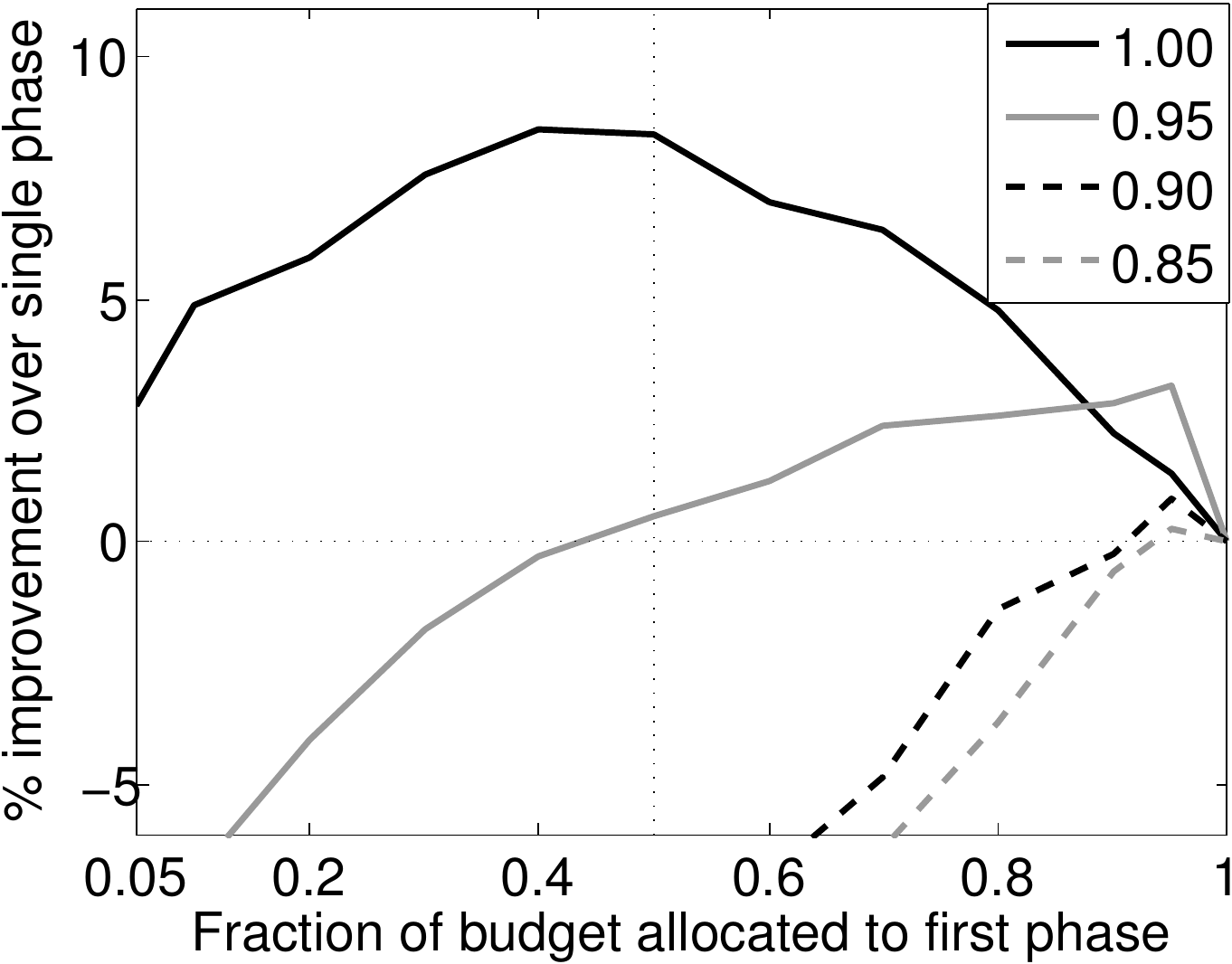} 
     \\ \centering (b)
\end{minipage}
   \caption{ 
  (a) Typical progression of diffusion for ${k=6}$ with different ${< k_1,d >}$ pairs (${k_2 = 6-k_1}$) on Les Miserables dataset 
   (WC model), 
(b) Typical observation of splitting budget ${k=200}$ (with optimal delay) for different ${\delta}$'s on 
High Energy Physics - Theory collaboration network
(WC model) 
}
   \label{fig:plots}
  \end{figure}

Since it is impractical to compute $S_2^{O(X,S_1,d,k_2)}$, consider 
$
f(S_1) =
\sum_X p(X) \sigma^X (S_1 \cup S_2 ^{G(X,S_1,d,k_2)}) 
$,
where $S_2 ^{G(X,S_1,d,k_2)}$ is a set of size $k_2$ obtained using greedy algorithm.
It can be shown that $f(\cdot)$ gives a $\left( 1-\frac{1}{e}-\epsilon \right)$ approximation to $g(\cdot)$,
where $\epsilon$ is small for large number of Monte-Carlo iterations while computing $f(\cdot)$.
Since greedy algorithm is not scalable, consider
$
h(S_1) =
\sum_X p(X) \sigma^X (S_1 \cup S_2 ^{W(X,S_1,d,k_2)}) 
$,
where $S_2 ^{W(X,S_1,d,k_2)}$ is a set of size $k_2$ obtained using 
generalized degree discount heuristic (GDD).
GDD
can be developed based on 
the argument for Theorem~2 in \cite{chen2009efficient}: until the budget is exhausted, iteratively select a node $v$ having the largest value of
$
\left( \prod_{x \in \mathcal{X}} (1-p_{xv}) \right) \left( 1+\sum_{y \in \mathcal{Y}} p_{vy} \right)
$,
where $\mathcal{X}=$ in-neighbors of $v$ already selected as seeds and $\mathcal{Y}=$ out-neighbors of $v$ not yet selected as seeds.
Using simulations, we observed for almost all $S,T$ pairs, that: 
\\
(a) $f(T)>f(S) \implies h(T)>h(S)$,
critical for set selection,\\
(b) $\frac{h(S)}{h(T)} \approx \frac{f(S)}{f(T)}$, 
critical for algorithms that depend on ratios of function values given by sets, e.g., fully adaptive cross entropy algorithm (FACE) with weighted update rule \cite{de2005tutorial}.

We now present a general algorithm for two-phase influence maximization.
Let $\mathcal{F}_1 (\cdot)$ and $\mathcal{F}_2 (\cdot)$ be objective functions for
the first and second phases, respectively.
Consider an algorithm $\mathbb{A}$ for single phase influence maximization.
\vspace{.1cm}
\hrule
\vspace{.1cm}
\textbf{Algorithm 1} Two-phase general algorithm (IC model)
\vspace{.1cm}
\hrule
\vspace{.1cm}
\begin{algorithmic}[1]
\renewcommand{\algorithmicrequire}{\textbf{Input:}}
\renewcommand{\algorithmicensure}{\textbf{Output:}}

\REQUIRE $G$, $k_1$, $k_2$, $d$

\STATE \textbf{First phase:} Find set of size $k_1$ using $\mathbb{A}$ for maximizing $\mathcal{F}_1 (\cdot)$ on $G$,
and
 run the IC model until time $d$ 

\STATE \textbf{Second phase:} At time $d$, construct $G^d$ from $G$ by deleting $\mathcal{A}^Y$;
assuming $\mathcal{R}^Y$ forms a partial seed set, find set of size $k_2$ using $\mathbb{A}$ for maximizing $\mathcal{F}_2 (\cdot)$ on $G^d$

\end{algorithmic}
\vspace{-.3cm}
\hrulefill

\noindent
We explore two special cases
(note that if $\mathbb{A}$ does not compute the expected spread, the two cases are identical):
\begin{nstabbing}
1.  Farsighted \= : \; $\mathcal{F}_1 (S_1) = h(S_1) \; , \;\;\; \mathcal{F}_2 (S_2) = \sigma(\mathcal{R}^Y \cup S_2)$ \\
2.   Myopic \> : \; $\mathcal{F}_1 (S_1) = \sigma(S_1) \; , \;\;\; \mathcal{F}_2 (S_2) = \sigma(\mathcal{R}^Y \cup S_2)$ 
\end{nstabbing}

\section{Experimental Findings}
\label{sec:simulations}

For studying diffusion using IC, we explore {\em weighted cascade (WC)} and {\em trivalency} models
\cite{chen2010scalable}.
Plots such as the ones in Figure~\ref{fig:plots}(a),
may help 
decide the ideal values of $k_1$ and $d$
based on the desired transient dynamics.
To capture the rate of diffusion,
we generalize $\sigma(\cdot)$ to
$
\sum_{t=0}^\infty \Gamma(t) \sigma^{(t)}(\cdot)
$,
where $\Gamma(\cdot) \leq 1$ is non-increasing,
and $\sigma^{(t)}(\cdot)$ is the expected number of recently influenced nodes at time $t$.
We consider $\Gamma(t)=\delta^t, \delta \in [0,1]$ in our experiments.
We discover FACE \cite{de2005tutorial} to be an effective method for concurrently optimizing over $k_1$, $d$, $S_1$,  by allowing each data sample to consist of a value of $k_1$ sampled from $\{1,\ldots,k\}$, a value of $d$ sampled from $\{1,\ldots,D\}$ ($D$ is some large delay after which, diffusion is guaranteed to stop), and a sampled set $S_1$ of size $k_1$.

For $\delta=1$, we observe that
$d=D$ (clearly) and $k_1 \approx k_2$ give best results (Figure~\ref{fig:plots}(b)), a reason being the trade-off between (i) the size of the observed diffusion and (ii) the exploitation based on the observed diffusion. 
For most values of $k$, the gain of two-phase diffusion over single phase one is 5-10\% for algorithms such as greedy, PMIA \cite{chen2010scalable}, FACE \cite{de2005tutorial}, and GDD, in absence of temporal constraints.
This gain is significant when the concern is monetary profits or a long-term customer base.
Also, myopic algorithms perform at par with farsighted,  while running a lot quicker (for greedy and FACE).
We conclude: (a) under strict temporal constraints, use single-phase diffusion, (b) under moderate temporal constraints, use two-phase diffusion with a short delay while allocating most of the budget to the first phase, (c) in absence of temporal constraints, use two-phase diffusion with a long enough delay with almost equal budget for the two phases.

\section{Future work}

There is a need for scalable algorithms that concurrently optimize over $k_1$, $d$, $S_1$
(perhaps exploiting unimodal nature of plots in Figure~\ref{fig:plots}(b)).
We considered a na\"ive, strict (exponential) decay function, which humbled two-phase diffusion for most $\delta$'s; a more realistic function needs to be studied.
One could study how multi-phase diffusion can be used to achieve a desired spread with a reduced budget.


\end{document}